\documentclass[journal]{IEEEtran}

\usepackage{cite}
\usepackage{url}
\usepackage{caption}
\usepackage{graphicx}
\usepackage{enumerate}
\usepackage{amsmath}
\usepackage{subcaption}
\usepackage{float}
\usepackage{stfloats}
\usepackage{multicol}
\usepackage{multirow}

\usepackage{dingbat} 
\usepackage{diagbox} 

\usepackage[linesnumbered,ruled,vlined]{algorithm2e}

\usepackage{balance}
\def\BibTeX{{\rm B\kern-.05em{\sc i\kern-.025em b}\kern-.08em
    T\kern-.1667em\lower.7ex\hbox{E}\kern-.125emX}}

\SetCommentSty{mycommfont}

\SetKwInput{KwInput}{Input}                
\SetKwInput{KwOutput}{Output}              

\usepackage[textsize=tiny, textwidth=1.5cm]{todonotes}

\usepackage[normalem]{ulem}

\newcommand{\snippet}[1]{``\texttt{#1}''}

\usepackage{tabularx, booktabs}
\newcolumntype{M}[1]{>{\centering\arraybackslash}m{#1}}

\usepackage[hidelinks,breaklinks]{hyperref}

\begin{document}
\bstctlcite{IEEEexample:BSTcontrol}

\title{A Privacy-preserving Mobile and Fog Computing Framework to Trace and Prevent COVID-19 Community Transmission}
%
%
%


\author{Md Whaiduzzaman,~\IEEEmembership{Member,~IEEE,}
        Md. Razon Hossain, Ahmedur Rahman Shovon,
        Shanto Roy,\\ Aron Laszka, 
        Rajkumar Buyya,~\IEEEmembership{Fellow,~IEEE,}
        and~Alistair~Barros
\thanks{Md Whaiduzzaman, and Alistair Barros are with Queensland University of Technology, Queensland, Australia (e-mail: wzaman@juniv.edu, alistair.barros@qut.edu.au).}
\thanks{Md. Razon Hossain, and Ahmedur Rahman Shovon are with  Jahangirnagar University, Dhaka, Bangladesh (e-mail: hossainmdrazon@gmail.com, shovon.sylhet@gmail.com).}
\thanks{Shanto Roy, and Aron Laszka are with University of Houston, TX, USA (e-mail: shantoroy@ieee.org, laszka.aron@gmail.com).}
\thanks{Rajkumar Buyya is with University of Melbourne, Australia (e-mail: rbuyya@unimelb.edu.au).}
}

%



\maketitle

\begin{abstract}
To slow down the spread of COVID-19, governments around the world are trying to identify infected people and to contain the virus by enforcing isolation and quarantine. However, it is 
difficult to trace people who came into contact with an infected person, which causes widespread community transmission and mass infection. To address this problem, we develop an e-government Privacy Preserving Mobile and Fog computing framework entitled PPMF that can trace infected and suspected cases nationwide. We use personal mobile devices with contact tracing app and two types of stationary fog nodes, named Automatic Risk Checkers (ARC) and Suspected User Data Uploader Node (SUDUN), to trace community transmission alongside maintaining user data privacy. 
Each user's mobile device receives a Unique Encrypted Reference Code (UERC) when registering on the central application. The mobile device and the central application both generate Rotational Unique Encrypted Reference Code (RUERC), which broadcasted using the Bluetooth Low Energy (BLE) technology.
The ARCs are placed at the entry points of buildings, which can immediately
detect if there are positive or suspected cases nearby. If any confirmed case is found, the ARCs broadcast pre-cautionary messages to nearby people without 
revealing the identity of the infected person. The SUDUNs are placed at the health centers that report test result to the central cloud application. The reported data is later used to map between infected and suspected cases. Therefore, using our proposed PPMF framework, governments can let organizations continue their economic activities without complete lockdown. 
\end{abstract}


\section{Introduction}\label{sec:intro}

\IEEEPARstart{T}{he} novel corona virus disease in 2019 (COVID-19) has spread rapidly
worldwide in a short duration. It caused a significant public health crisis worldwide, and by the end of May 2020 (i.e., within the five months of its first infection detection), over $6.16$ million persons were infected, and over $372$ thousand have died~\cite{HomeJohn79:online}.
Therefore, governments worldwide seek solutions to minimize the infected cases from the COVID-19 pandemic by employing mobile application based contact tracing~\cite{sharma2020use,chan2020pact}. Mobile apps can help trace both infected and suspected cases in almost real-time, and governments are rushing towards developing and deploying such applications and frameworks. However, several applications raise significant privacy issues as they collect sensitive and personally-identifiable data from users, and lack user control and transparency in data processing or usage~\cite{sharma2020use,cho2020contact,chan2020pact,raskar2020apps}.

Governments are enforcing temporary lockdowns of cities to slow down the spread of COVID-19, causing tremendous economic losses. However, we can alleviate economic impact by avoiding wide-scale lockdowns and performing more targeted isolation.
Therefore, introducing fog computing in economic zones (e.g., shopping malls, organization buildings) can ensure continued economic activities by alerting nearby people while mobile computing (mobile apps) can help to trace the infected and suspected cases. However, to the best of our knowledge, there is 
no integrated fog computing framework alongside contact tracing mobile apps that allows tracing community transmission while preserving users' data privacy. Therefore, we introduce the following research questions to find an appropriate solution to the cause.

\begin{itemize}
    \item[\textbf{Q1.}] \textbf{Background and Issues:} What are the issues and privacy concerns in existing contact tracing apps?
    \item[\textbf{Q2.}] \textbf{Mobile and Fog Computing:} How to utilize mobile and fog computing to trace and prevent COVID-19 community transmission?
    \item[\textbf{Q3.}] \textbf{Privacy-preserving Framework:} How to develop an automated privacy-preserving e-government framework?
\end{itemize}

We answer the first question by looking into the background and issues of existing application frameworks as well as user data privacy concerns (Section~\ref{sec:background}). We find that there are several mobile application frameworks developed by governments and third parties to trace the COVID-19 community transmission. However, most of these applications and frameworks have failed to ensure user data privacy and suffer from other issues, such as mandatory use of apps, excessive data gathering, questionable transparency of source codes and data flow, unnecessary data usage or processing, and lack of user control in data deletion.

\begin{figure*}
    \centering
    \includegraphics[width=.7\textwidth]{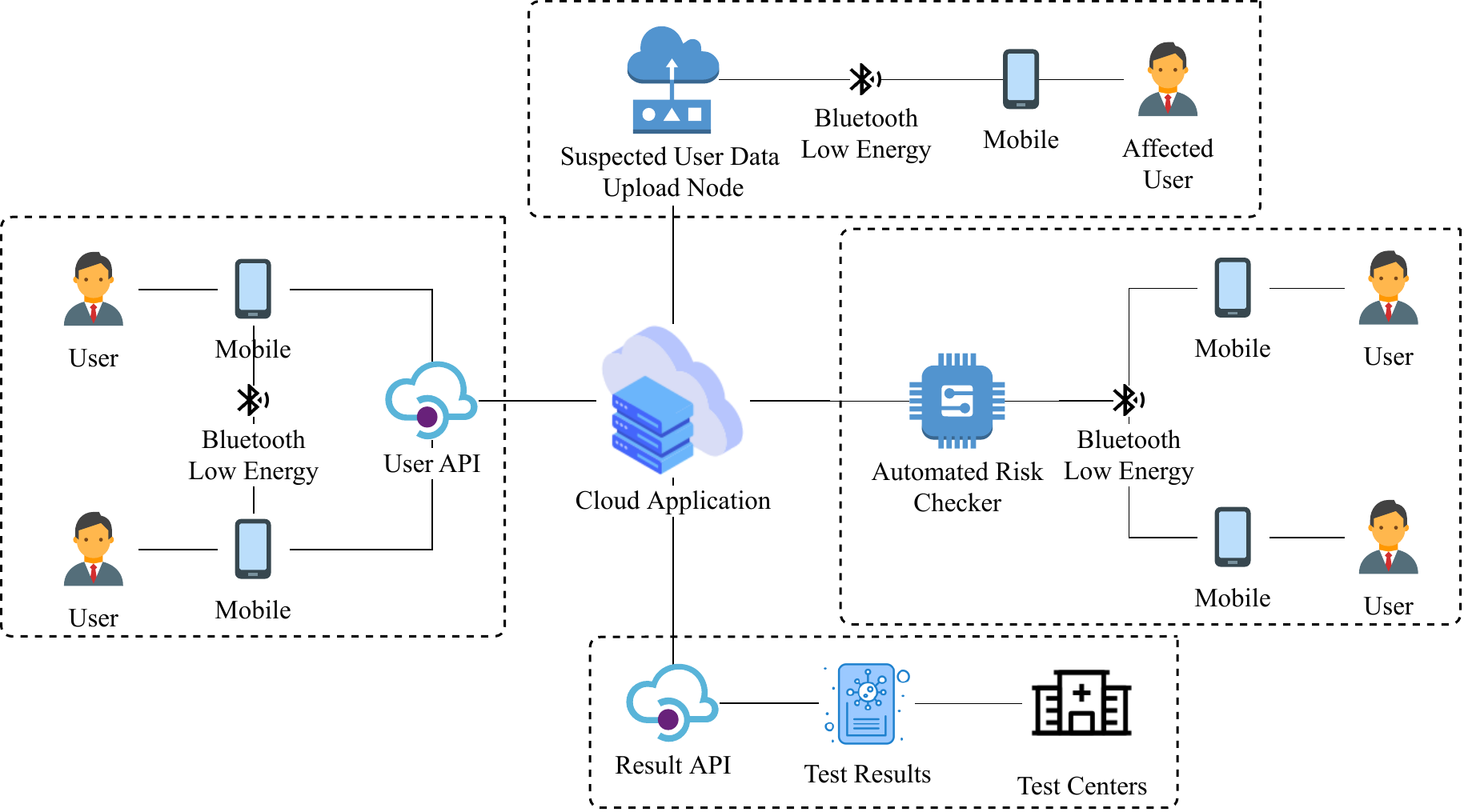}
    \caption{System Overview}
    \label{fig:block}
\end{figure*}

We answer the second question by presenting the design considerations, architecture, and workflow of our e-government application framework that utilizes mobile and fog computing (Section~\ref{sec:method}). The system consists of two types of fog nodes (ARC and 
SUDUN), several RESTful APIs, and a central application. Users can register themselves using User API. The ARC is used to check the risk of users visiting any public places (e.g., shopping mall, organization building). Test centers send the COVID-19 test results using Result API to the central application. If the test result is positive, then the user is requested to upload locally stored contact tracing information to the cloud using SUDUN or directly from mobile app. Figure~\ref{fig:block} presents an overview of the proposed system.


We answer the third question by discussing the implementation overview and user data privacy solutions in our framework (Section~\ref{sec:development}). Here, we discuss the framework development based on the Amazon Web Services (AWS) solutions. Then we present the privacy preservation based on user control (voluntary, compliance, and user consent), minimal data collection (mobile number, postal code, and age group), data destruction at user's will, transparency (open source codes, clean data flow), and limited further usage of data.


    

\paragraph*{\textbf{Scope}}
We intend our framework to be 
deployed and controlled by the central government of a country. 
Governments have access to the test results and can control such an integrated mobile-fog computing framework. 
Moreover, relying on a private entity
to manage such a framework can limit the preservation of data privacy. 
Additionally, in this work, we primarily consider and focus on user data privacy issues. 
Regardless of building a standard and secure data processing framework, we do not discuss 
advanced security threats related to mobile, fog, and cloud layer as there are rich literature on existing security measures~\cite{roman2018mobile,yi2015security,singh2016survey}.

\paragraph*{\textbf{Outline}}
The rest of the paper is organized as follows: Section~\ref{sec:background} introduces the privacy and general issues of existing contact tracing frameworks. Section~\ref{sec:method} discusses the design considerations, system components, and workflow of our proposed e-government framework. Section~\ref{sec:development} presents the implementation overview and privacy-oriented solutions of the framework. Finally, Section~\ref{sec:related} discusses a few related privacy-preserving frameworks followed by a conclusion.
\section{Background}\label{sec:background}
Contact tracing and isolation of cases are required to control infectious disease outbreaks, and the consequence depends mainly on government actions and citizen responses~\cite{hellewell2020feasibility}. Manual tracing is difficult and time-consuming in a pandemic situation, and governments are pushing for digital surveillance to contain the spread of the virus in the context of health informatics~\cite{BigDataHealth}.
However, these mobile apps have raised questions over user data privacy since digital surveillance involves location tracking, limits individual freedom, and expose confidential data~\cite{raskar2020apps}.

\subsection{Contact Tracing}





Governments can stop transmission of COVID-19 if they identify cases and their contacts quickly and get them to limit their connections with other people. Cases should isolate themselves as long as they are infectious- for at least $10$ days after they became ill. Contacts must quarantine for $14$ days after the last contact with an infectious patient~\cite{pung2020investigation}.
Some cases may have close contact with many people because of where they have been or where they live, and these situations should immediately be reported.

Contact tracing might be difficult as an infected case may not remember all people who came in their contacts. Additionally, an infected person may not know or remember the phone numbers and address of their contacts. It may take longer for authorities as well due to the required time to identify and get in touch with contacts. Therefore, mobile-based contact tracing might help track several contacts and determine who is at highest risk for infection. 

Contact tracing usually requires three primary steps: contact identification, listing, and follow-up~\cite{raskar2020apps}. People with contact with an infected person are considered as suspected cases if they were in proximal range for more than 15 minutes, within 6 feet~\cite{CDC:online} or the distance of more than 6 feet, but stayed nearby for an hour~\cite{COVID19C58:online}.






\subsection{Contact Tracing Frameworks and Applications}
Due to the necessity of tracing infected or suspected cases, governments around the world have developed several tracing applications and frameworks~\cite{cho2020contact}. The most common features of these applications are live maps and news updates of confirmed cases, location-based tracking and alerts, quarantine and isolation monitoring, direct or indirect reporting, self-assessment, and COVID-19 education~\cite{sharma2020use}. Some governments involved third parties to develop such applications and encouraged citizens to use these applications. 



\subsubsection{Privacy Concerns}
Since governments have been rushing to build tracing applications, the least they have considered about user data privacy. In most cases, users can be monitored and tracked in real-time without user’s consent. If such mobile applications store the location history, user movement can be traced as well~\cite{berke2020assessing}. Apart from location tracking, there are several other user privacy issues, such as excessive data collection, obscure data flow, lack of user control, and data usage policies.

Abeler et al.\ suggested that we can achieve contact tracing and data protection at the same time by minimizing data processing in the existing frameworks~\cite{abeler2020covid}.
However, many applications are not following such solutions and Howell et al.\ has suggested five primary privacy concerns for COVID-19 application frameworks~\cite{Afloodof14:online}.
    \begin{itemize}
        \item \textit{Voluntary or Mandatory:}
        It should be a voluntary act whether users download and use such tracing apps. With the growing concern over data privacy, unnecessary data collection, location tracking, and other issues, users must have free wills to decide. The government or any third party cannot mandate users to use these apps in any circumstances.
        \item \textit{Data Usage Limitation:}
        People are concerned over the collected data usage for personal safety reasons~\cite{simko2020covid}. Therefore, the collected data must have usage limitations. For example, tracing data can only be used for public health and safety. Traced data cannot be used for any other purpose, e.g., law enforcement.
        \item \textit{Data Destruction:}
        Mobile applications or a framework should automatically delete user records after a particular period~\cite{dubov2020value} (e.g., usually 14-21 days and no longer than 30 days). Otherwise, users should have manual control over data deletion from the app or the central server.
        \item \textit{Minimal Data Collection:}
        Several applications collect excessive, unnecessary data from its users, for example, an application named
        \snippet{Aarogya Setu} requires name, phone number, age, gender, profession, and details of countries visited in the last 30 days.
        Also, Geo-location tracing is unnecessary alongside Bluetooth or other similar wireless technologies~\cite{Microsof7:online}.
        
        \item \textit{Transparency:}
        The entire process of data collection and usage should be transparent to preserve user privacy. Application frameworks should have publicly available policies, clear and concise data flow and database, and open-source codes for transparency. Additionally, users should have full control over their data usage. Therefore, developers must follow the \emph{compliance} and \emph{consent} rules (GDPR, HIPAA, CCPA, etc.) strictly~\cite{GDPRCompliance,dubov2020value}.
        
    \end{itemize}

Apart from the privacy issues mentioned above, there are certainly other things to consider, as well. For example,
the mobile application generated IDs can be breached, decrypted, and resulted in exposing user information. Additionally, applications controlled by the third party may pose a severe threat as they can misuse the collected user data. Therefore, in terms of privacy, user control over the owner's data and transparency are notable factors.



\vspace{2mm}
\subsubsection{General Concerns}
Apart from several data privacy issues, different design issues are prevailing in the contact tracing apps. Many applications require a constant internet connection while it is entirely unnecessary. In some other implementations, the user can not turn off the background service of the apps, and these apps do not feature turn-off option.
The apps continue to work at random times, such as while staying at home or sleeping. Therefore, it causes battery drains too fast.

Tracing correctness depends on the distance and period of contact. Several applications stores the RUERC of a nearby user device even there is a wall between two persons. Moreover, rushing to develop such applications result in \emph{false positive} suspected victim considerations and community transmission. Communication between multiple platforms may appear troublesome and can lead to unexpected behavior. Therefore, using Google/Apple API might be helpful to introduce Bluetooth communication between two devices of different platforms (Google Android and Apple iOS). Apart from these issues, \emph{insecure source code}, \emph{weak data flow and process}, and different \emph{Bluetooth-based device attacks} can cause security hazards.

\subsubsection{Google/Apple Exposure Notification API}
Google and Apple announced their privacy-preserving exposure notification in April 2020 and released phase one in May 2020~\cite{ExposureFAQ42:online}. The API uses BLE technology and applies different hashing algorithms to generate different keys on a specific time interval~\cite{ExposureBS20:online} to prevent wireless tracking. Infected cases upload only the daily generated keys of the past 14 days, and other users download lists of these keys of infected cases of their corresponding region. 

All the key-generation and risk-level checks are performed in the user's mobile device, preserving user privacy. However, this might also be a concern about the resource consumption of that particular mobile device~\cite{HashEC}. Moreover, as the user is only uploading their keys, the authority can not notify the contacts of the infected case immediately. Some other essential functionalities, such as preventing community transmission and identifying the asymptomatic spreaders, can also be troublesome.

The responsibility of implementing this technology is on the corresponding public health authorities. However, the authority is required to follow the guidelines of privacy, security, and data control rules as well as the development criteria such as file and data format of storing, uploading, and downloading keys~\cite{ExposureFAQ42:online,ExposureTerm72:online}.
This collaborative development of the Exposure Notification ensures that both the platform (Android and iOS) transmits and receive similar keys, and the risk level the app calculates is also similar.

Current contact tracing apps backed by the governments have numerous design and privacy issues due to the rush for community transmission tracing. Regardless of addressing existing issues, the question remains if the governments can continue the economic activities alongside. Therefore, our integrated mobile and fog computing-based framework can solve the issues by effectively tracing community transmission while organizations can run their economic activities. 
\section{Framework Design}\label{sec:method}
Our proposed privacy-preserving e-government framework has four major components: user mobile device and two types of fog nodes (ARC and SUDUN), and a central cloud application that integrates these nodes. Mobile Unit consists of BLE, privacy dashboard, filter algorithm, file storage, and communication service. A user mobile device advertises it's own RUERC, scans for RUERCs of other nearby devices, and save the filtered RUERCs (see filtering Algorithm~\ref{algo:suspect_filter}) in the file storage. 

\begin{figure*}[!ht]
    \centering
    \includegraphics[width=.78\textwidth]{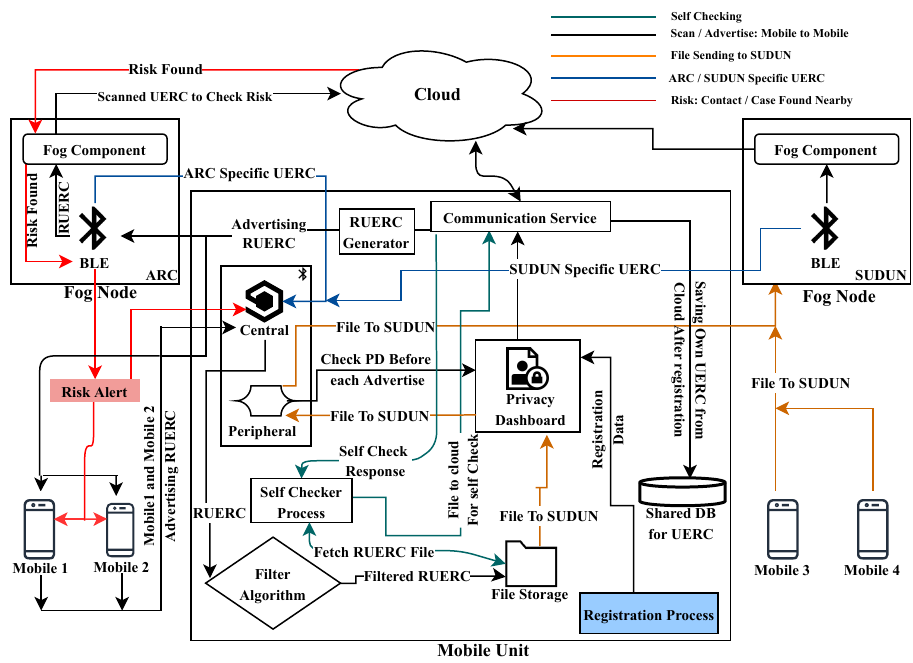}
    \caption{Mobile and Fog Computing Framework for Tracing and Preventing Community Transmission}
    \label{fig:mobile_fog_framework}
\end{figure*}

Fog-based IoT-healthcare provides optimization of data communication, low power-consumption, and improves efficiency in terms of cost, network delay, and energy usage~\cite{mahmud2018cloud,IoTCloud}. 
The BLE in the ARC and SUDUN advertises a specific predefined UERC, and the mobile unit does not save these UERCs to file storage. When the user is in proximity to ARC (hospital, shopping malls, office), BLE of the ARC receives all the RUERCs around, and the fog component checks if there is any infected or suspected case. The cloud application responds with a positive or negative result without disclosing the victim's identity. If there is a positive case, the ARC transmits another predefined UERC that signifies the risk level and alerts all the mobile units around, including the infected or suspected case, hence preserving the victim's privacy. The other fog node, SUDUN, is set either in the test centers or where the authority finds it necessary. When the mobile unit receives the predefined UERC from this fog node, it checks its privacy dashboard. If the privacy dashboard allows the mobile unit to send the file to SUDUN, the mobile unit establishes a connection with SUDUN and send the file. Then, the fog component in SUDUN transfers the file to the corresponding cloud.
Here, 
Figure~\ref{fig:mobile_fog_framework} presents the detailed workflow our proposed integrated mobile and fog computing framework.

\subsection{System Features}
We introduce the following features of mobile application and fog nodes in our integrated framework:

\begin{enumerate}
    \item \textit{Contact Tracing:} An user gets a hashed unique reference code upon registration from the system. Every two hours, another unique reference code is generated in the application, which is being shared with nearby devices upon user consent. It ensures that the broadcast data cannot be used to trace an individual. The same hashed value will be generated in the cloud application, and it can be used to check the risk level of any individual without disclosing his or her identity with others. 
    \item \textit{Self-checking:} Users can check if he or she was nearby any infected victim in the last 14 days using the mobile application. The application will upload the locally stored reference codes of the device it came in contact in the last 14 days. The cloud application will check if any of the uploaded RUERC is listed as an infected victim or suspected victim. The application will notify the user accordingly.
    \item \textit{Minimum Mobile Computation:} 
    Our proposed framework ensures minimum computation by enabling user control to turn on or off scanning, and background services. Apart from that, the mobile application features delay broadcast by avoiding unnecessary frequency and requires minimum internet connection as users only need internet while registering and uploading data for self-check.
    \item \textit{User Data Privacy:} User personal data is not stored or shared in the system. We ensure minimum data collection and avoid user location tracking or digital surveillance. Fog nodes do not identify any infected victim. 
    \item \textit{Fog Node Alerts:} 
    To minimize community transmission, we introduce automatic risk checker in public places such as shopping malls and office buildings. As there is a chance of revealing the user identity, we introduce \emph{time delay} and \emph{minimum entry} of people before broadcasting simple alert messages. These messages only request users to take precautions; do not reveal any RUERC or risk radius.

\end{enumerate}

\subsection{System Components}
Our framework has four
major components: a mobile application that broadcasts its RUERC, and stores received RUERCs from nearby devices. The ARC checks for infected cases and broadcast alerts in organization building. The SUDUN uploads data from the device of infected cases. Finally, a central cloud application integrates all these mobile and fog nodes and manages collected data. 

\subsubsection{Mobile Application}
Users download the application from the google play store or govt server and install it in the mobile device. The application broadcasts its RUERC and receives RUERC of other devices around. It also detects the special predefined UERC broadcasted by Automatic Risk Checker (ARC) and alerts the user about an infected or suspected case around.

\subsubsection{Automatic Risk checker (IoT/Fog) (ARC)}
This fog device is set inside the hospital, shopping mall, educational institutes, and in all government-supervised organizations where there is a possibility of community transmission. The fog can receive the RUERC of the mobile devices within the Bluetooth range and interacts with cloud service to detect any infected or suspected victim. If any RUERC is found as an infected or suspected victim, it broadcasts a specific UERC throughout the place. The mobile application near the fog receives this specific UERC and alerts the user about the risk. As the fog node is registered via cloud fleet management service, the ARC node can be monitored in real-time. Thus it allows identifying any ARC node with a high rate of COVID-19 affected patient's presence. If the number of infected patients or the number of suspected patients gets higher than a predefined maximum threshold value per hour, it automatically generates emergency alert service to the organization's authority by sending emails and messages. It informs the nearby health authorities as well. The notification is broadcast only when there are at least $5$ persons within the range of ARC to ensure the privacy of the infected victim. On the contrary, the notification is not sent instantly when the ARC identifies an infected or suspected victim. It notifies the users after a certain period, which ensures the identity of the infected or suspected victim is not disclosed to others.

\subsubsection{Suspected User Data Uploader Node (IoT/Fog) (SUDUN)}
These fog nodes are similar to ARC and are set at the health centers or the COVID-19 test centers. When a test center reports a positive test result, the infected person can voluntarily allow the application to enable uploading its RUERC list from the privacy dashboard and keep the mobile phone within the range of a SUDUN.
A SUDUN automatically connects with the user's device and fetches the list of stored lists of hashed RUERC from the device. Here, we enable monitoring each SUDUN using a cloud dashboard on a real-time basis. Additionally, the infected victim can also upload the RUERC list by using the mobile application and an internet connection without the help of a SUDUN.

\subsubsection{Cloud/Server}
The government provided secure data storage and computing server. Receiving and storing data from SUDUN, mining to predict essential patterns or super spreaders, computing to provide required information to ARC, and handling standard authentication to maintain security and privacy are its primary responsibilities.

    
    

\subsection{System Workflow}
\subsubsection{User Registration}
\begin{figure}[!ht]
    \centering
    \includegraphics[width=.5\textwidth]{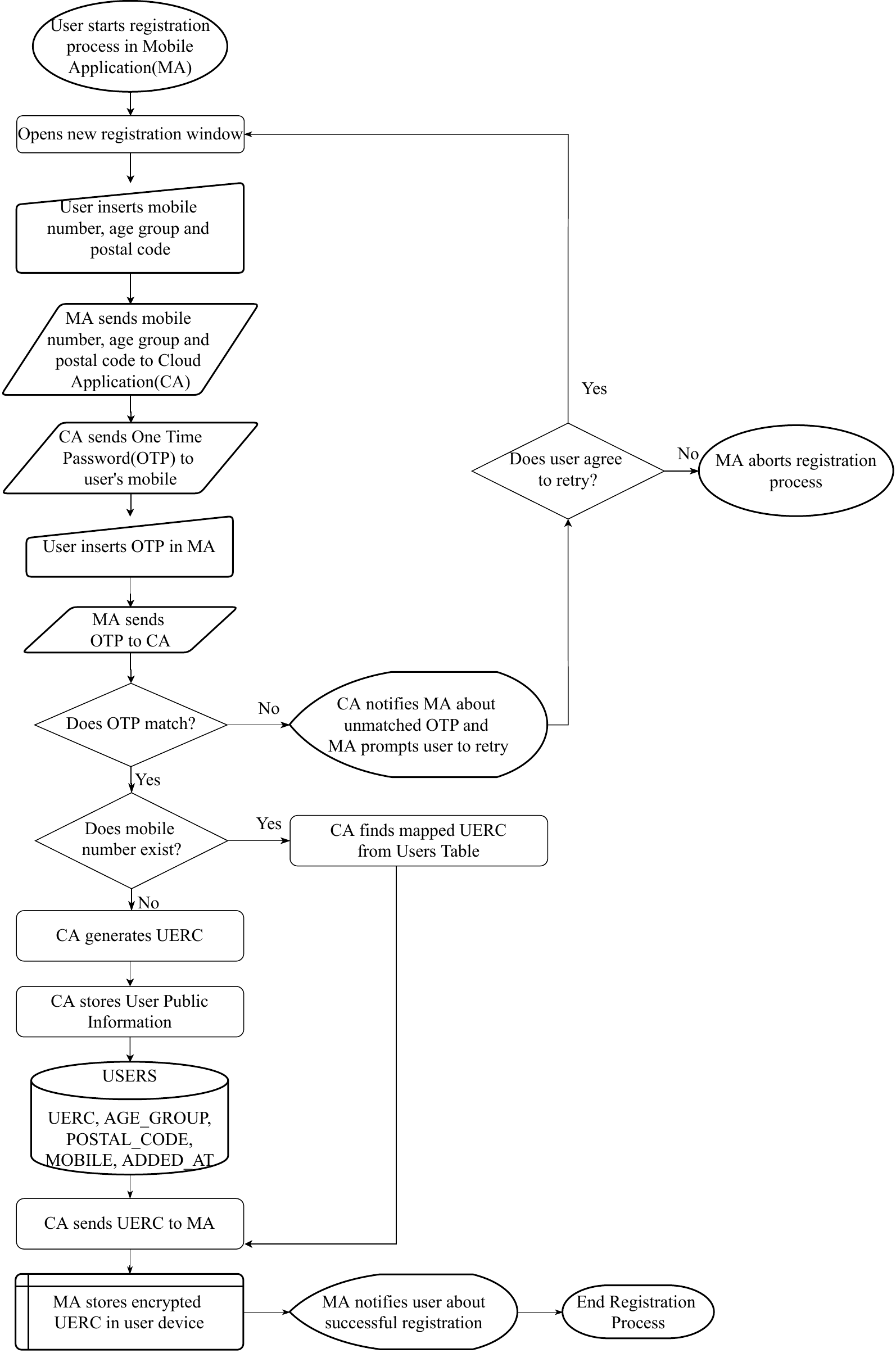}
    \caption{User Registration Process}
    \label{fig:registration}
\end{figure}

Initially, users need to register in the cloud using mobile number, age group, and postal code. The mobile number field is mandatory; however, the age group and postal code are optional. The Cloud Application generates and sends a One Time Password(OTP) to the user-submitted mobile number through SMS. Then the user enters the OTP in the mobile application, and the app sends the OTP to Cloud. The cloud application matches the user sent OTP, with the generated one. If both matches, the cloud application generates a UERC for the user and send it back to the user installed mobile app. 
Finally, the mobile app stores the UERC in encrypted format in the user device and then proceeds to scan IDs broadcasted by other users. The user device itself broadcasts a rotational UERC (derived from the initial UERC) as well. Figure~\ref{fig:registration} presents the complete registration process.


\subsubsection{Key Generation}
\begin{figure}[!t]
    \centering
    \includegraphics[width=.5\textwidth]{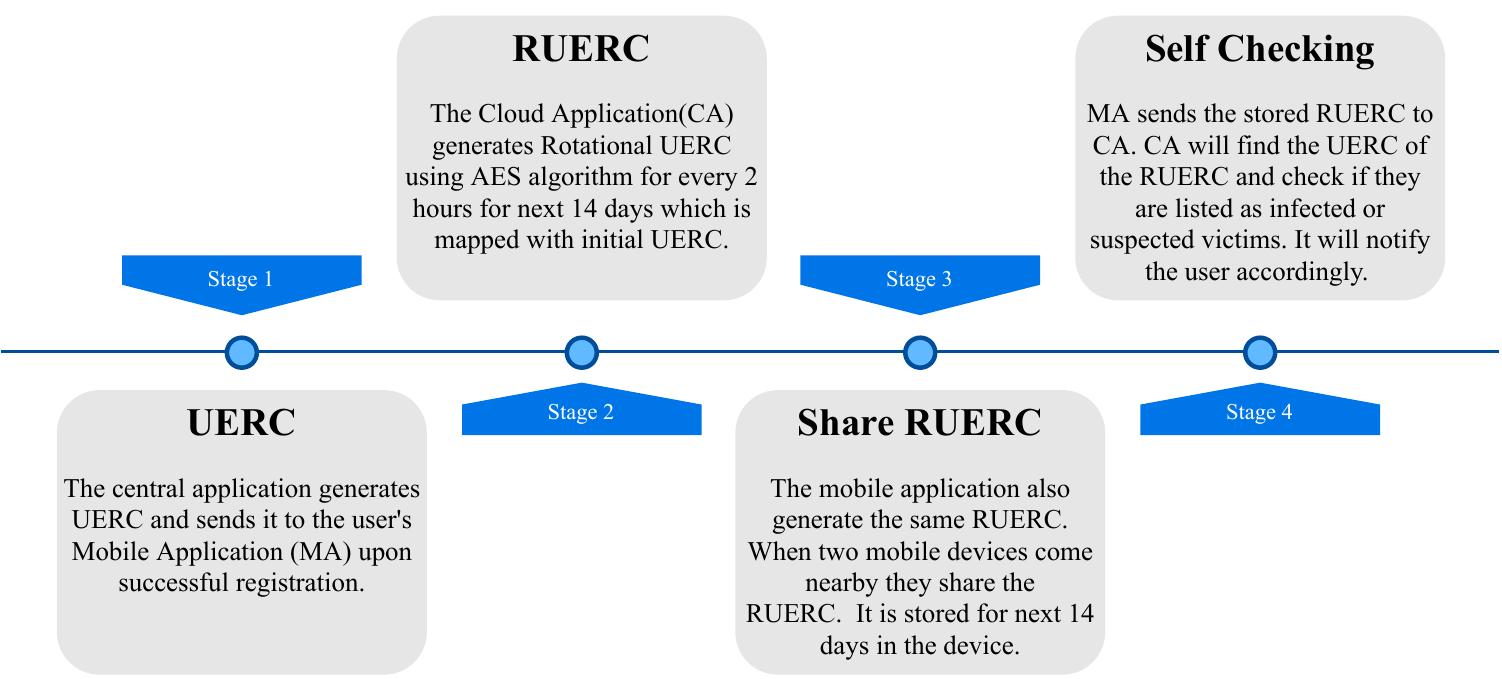}
    \caption{Key Generation Steps}
    \label{fig:key_gen_steps}
\end{figure}

Our framework generates Rotational UERC (RUERC) with an interval of two hours. A person is considered a suspect case if he stays nearby to an infected person for an hour. As the RUERC changes every two hours, possibly the users are supposed to receive more than one RUERC if they stay nearby longer than an hour (e.g., in case they are neighbors). Our mobile application ensures that even if a person comes in close at any minute, it will store individual key and compare against the timestamp. It increases the accuracy of finding contact without a rigorous need for computation. Moreover, the RUERC enhances the privacy of the user because always broadcasting the same UERC may result in wireless tracking.
While registering, the mobile unit receives a UERC from the cloud, and from this UERC, it generates RUERCs using the AES algorithm on each two-hour interval for the next 14 days. When two devices come close, they transmit and save these RUERCs. Figure~\ref{fig:key_gen_steps} presents the steps of RUERC generation.

\subsubsection{RUERC Scan Service}
Scan service allows the application to run in the background. It creates a new thread so that the user does not feel any obstacle while using other mobile apps. 
The user has the privilege to stop and then restart running the service at any time. Scanning and advertising RUERC, creating files, and saving data are its primary responsibilities.

Scan service uses Bluetooth Low Energy (BLE) technology to share RUERC between two mobile devices. 
The BLE consumes significantly less power to communicate, control, and monitor IoT devices. It uses \snippet{central} and \snippet{peripherals}, which define a network called piconet \cite{ble1}. Central scans for the advertisement and peripheral makes the advertisement. In our framework, the advertisement consists of the defined RUERC.
The Generic Attribute Profile (GATT) is designed to send and receive short pieces of data known as "attributes," and it is built on Attribute Protocol (ATT), which uses 128-bit unique ID~\cite{ble1}. In our framework, to maintain the minimal data collection policy and preserve device power, we are omitting the use of GATT and ATT. We are using emph{peripheral} to transmit the RUERC and the \emph{central} to scan these RUERCs. While scanning, we filter these RUERCs with our \emph{suspect filtering algorithm} and save accordingly.

\subsubsection{RSSI}
The framework uses a Received Signal Strength Indicator (RSSI) to identify the distance between devices. RSSI shows the strength of the received signal. It is calculated in dB and depends on the power and chipset of the broadcasting device. It also depends on the transmitting medium. If there is any obstacle between the receiver and the sender, the signal strength will decrease, so the RSSI. Therefore, for a different manufacturer, the value of RSSI in a specific medium shows different values. However, for a particular manufacturer, RSSI shows the intensity of the signal strength. 
RSSI does not provide the accurate distance but using path loss model \cite{kumar2009distance}, an approximate distance can be found,
\begin{equation*}
    RSSI = -10 \cdot n \cdot \log_{10}(d) + C
\end{equation*}
Here, \textit{n} is the path loss exponent that depends on the transmitting medium, \textit{d} is the distance, and \textit{C} is a constant.


\subsubsection{RUERC Storage in mobile device}
To maintain user privacy, the mobile application creates two files in the internal storage of the mobile device. Even the user does not have permission to access these files. When the peripheral receives any signal, one file temporarily stores this signal information, and other stores the information of the filtered signal. When the device comes within the range of SUDUN, the signal information of the filtered file is sent to SUDUN or directly to the cloud to perform self-check.

\subsubsection{Suspect Filtering}
According to CDC, a distance of six feet between the case and contact is safe to maintain~\cite{CDC:online}. A person is considered a contact if he/she is within six feet for at least fifteen minutes or is within the proximity for an hour~\cite{COVID19C58:online}.
The scan service scans all the RUERCs within the range of Bluetooth. However, considering the rules mentioned above, continuous scanning is unnecessary as it causes battery and memory consumption. Therefore, the scanning service scans for $700 ms$ with three minutes interval. We develop our suspect filtering algorithm considering these factors so that the power and memory consumption is minimal, and identifying suspects is more accurate.

There are two types of files in the mobile device. One contains all the signal information (RUERC, distance, and timestamp) it receives from nearby devices on each interval. If any received signal passes the filtering conditions, associate information (RUERC, distance, timestamp, and duration) is stored in another file. This file is called the \emph{final file} and sent to the cloud once the user is found COVID-19 positive. A specific RUERC is stored in the final file only once a day. Here, the suspect filtering algorithm is shown in Algorithm~\ref{algo:suspect_filter}.


\begin{algorithm}[!ht]
\DontPrintSemicolon
  
\tcc{
$T_F$ = Temporary File, $F_F$ = Final File,\\
$T_c$ = Current Time, $T_r$ = Registerd Time,\\
$T_{15}$ = 15 minutes, $T_{60}$ = 60 Minutes,\\
$D_c$ = Current Distance, $D_r$ = Registered Distance, 
$D$ = Standard Distance (6 Feet), $C_t$ = Contact, $C_s$ = Case,
$S_{c}S_{n}$ = Scanned Signal, $R_{c}S_{n}$ = Recorded Signal in $T_F$,\\
$ASU$ = ARC specific UERC, $RSU$ = Risk specific UERC, \\
$SSU$ = SUDUN specific UERC
}

\ForEach{Signal in $R_{c}S_{n}$}    
        { 
        	 \If{$S_{c}S_{n}$.contains(Signal) == false}
                {$R_{c}S_{n}$.remove(Signal)}
        }
        
\ForEach{Signal in $S_{c}S_{n}$}    
        { 
        	 \If{Signal.RUERC == self.RUERC OR Signal.RUERC == $ASU$}
        	    {
        	        return
        	    }
        	 \If{Signal.RUERC == $RSU$}
        	    {
        	        showRiskAlert()
        	        return
        	    }
        	 \If{Signal.RUERC == $SSU$}
        	    {
        	        sendFileToSUDUN()
        	        return
        	    }
        	 \If{$R_{c}S_{n}$.contains(Signal) == false}
                {$R_{c}S_{n}$.add(Signal)}
            \Else
            {
            	Set $T_c$ = Signal.time, $T_r$ = $R_{c}S_{n}$.get(Signal.RUERC).time, $D_c$ = Signal.distance, $D_r$ = $R_{c}S_{n}$.get(Signal.RUERC).distance\;
            	
            	\If{$D_c <= D$ AND $D_r <= D$}
                {
                    \If{$T_c - T_r >= T_{15}$}
                    {
                    $C_t$ = Signal\;
                    \If{($F_F$.contains($C_t$) AND $F_F.get(C_t).time.date == C_t.time.date$) OR isUserConsentToStoreData == false}
                    {
                        return
                    }
				    $F_F$.add($C_t$)\;
				    }
				    
				    \Else
                    {
                    	$D_r$ = $D_c$\;
                    }
                }
                
                \ElseIf{$T_c - T_r >= T_{60}$}
                {
                	$C_t$ = Signal\;
                	\If{($F_F$.contains($C_t$) AND $F_F.get(C_t).time.date == C_t.time.date$) OR isUserConsentToStoreData == false}
                    {
                        return
                    }
			        $F_F$.add($C_t$)\;
                }
                
                \Else
                    {
                    	$D_r = D_c$\;
                    }
            }
        }

\caption{Filtering suspected RUERC in mobile application}
\label{algo:suspect_filter}
\end{algorithm}

\subsection{Data Processing}
\begin{figure}[!ht]
    \centering
    \includegraphics[width=0.49\textwidth]{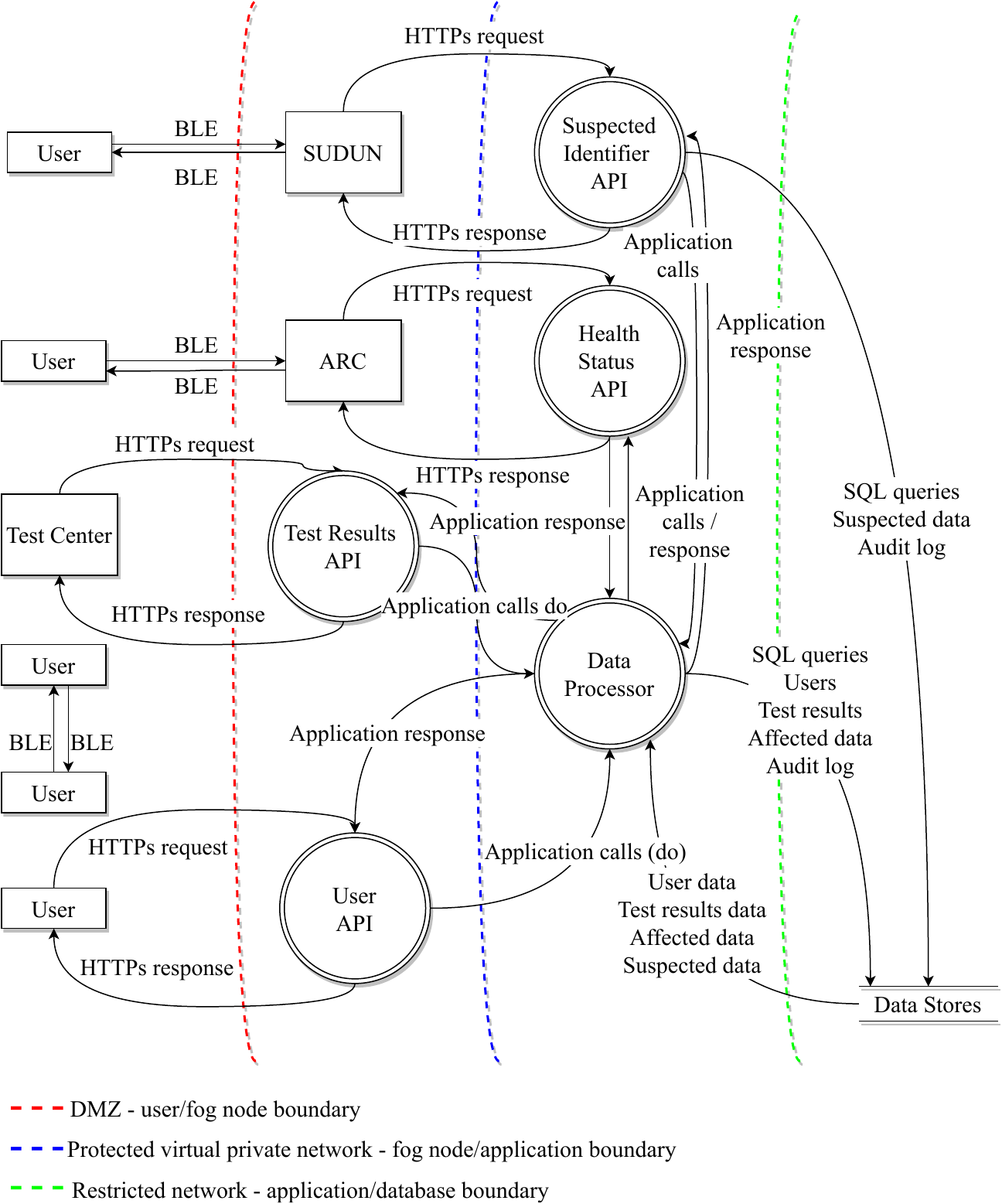}
    \caption{Process Flow Diagram with Privacy and Security Boundaries}
    \label{fig:process_flow_diagram}
\end{figure}

The framework is divided into three data processing layers to preserve user privacy and restrict access to data in various processes. Users can register to the system using a public communication process via mobile application. In this same access layer, the test centers can send the test results using RESTful API services through HTTPs protocol. The users can connect ARC and SUDUN fog node through the BLE protocol and send data to fog nodes. The computation and initial screening of user-uploaded data in fog nodes is done in a protected network layer. The APIs which are consumed by user mobile applications, test centers, and fog nodes are also in this layer. The main data processor consisting of a set of application nodes resides in a protected network, and users cannot access the data processor directly. The processed data is stored in data storage, which is configured in a restricted private subnet. This subnet can only be accessed from the ancestor private subnet, not from any public subnet. The data layer is protected from SQL injection as it is not directly connected through any user entries. Figure~\ref{fig:process_flow_diagram} displays the process flow diagram.

\section{Framework Development} \label{sec:development}
In this section, we discuss the implementation overview of our framework with regards to \emph{Amazon Web Service}. Additionally, we discuss the database design, analysis of our contact tracing graph, and our privacy-preserving solutions.

\subsection{Implementation Overview}
We have implemented our framework using Amazon Web Services (AWS)~\cite{robinson2008amazon,AmazonWe40:online}. As it is a generic framework, it can be applied in other IoT/Cloud platforms too. The fog nodes, ARC and SUDUN, consist of AWS Greengrass components and Lambda functions. They are connected to the AWS IoT Core service using the MQ Telemetry Transport (MQTT) protocol. The authenticity and security of the fog nodes are ensured by using IoT Device Defender. IoT Device Management service is used to monitor and audit the fog nodes. Incoming data from the fog nodes are passed to Simple Queue Service(SQS). The queue data is processed via Lambda functions and moved into the application containers, which is managed via the Kubernetes cluster within an auto-scaling group for dynamic scaling of the master nodes and worker nodes.

\begin{figure*}[!ht]
    \centering
    \includegraphics[width=.87\textwidth]{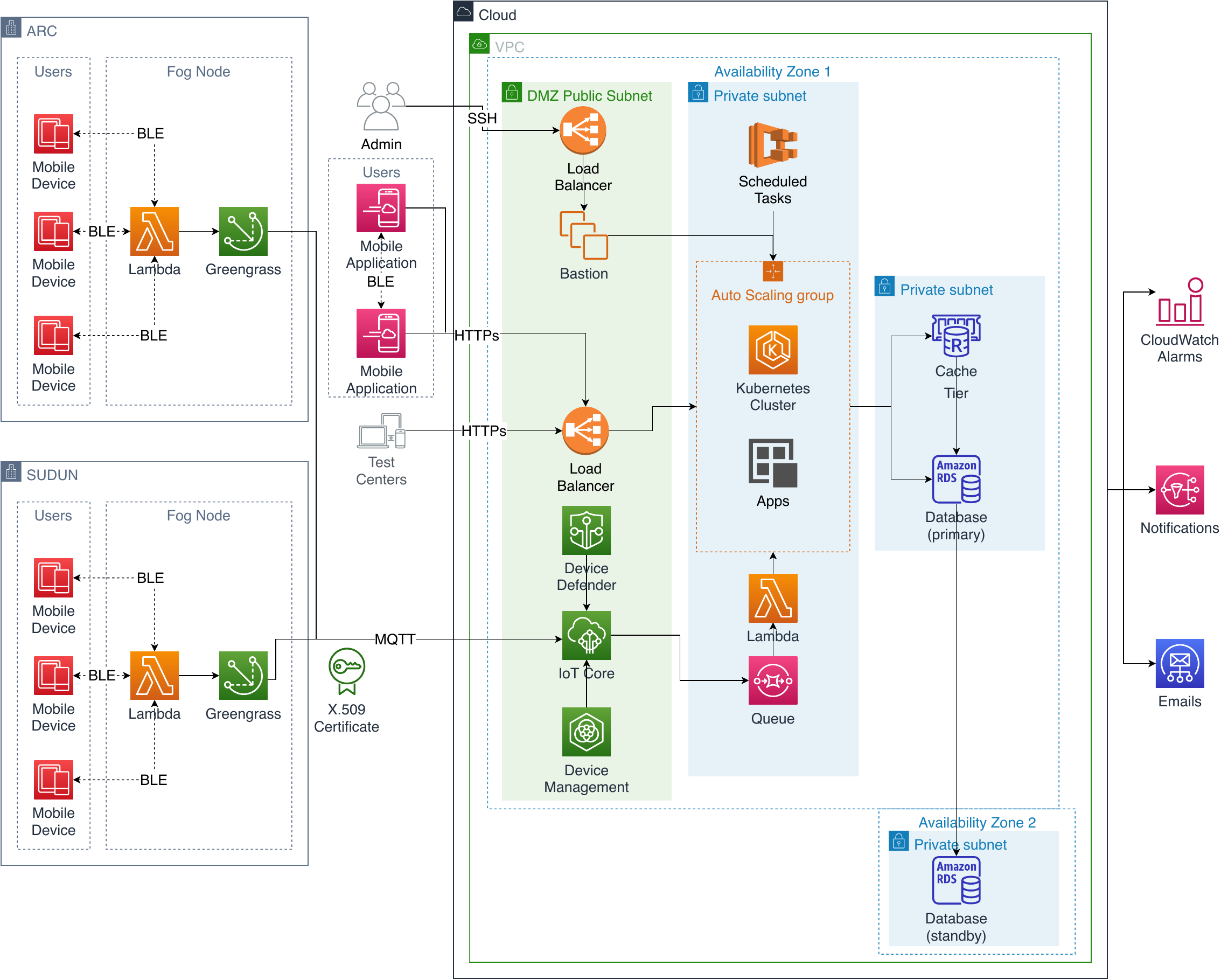}
    \caption{Implementation Overview with regards to AWS}
    \label{fig:Implementation Overview}
\end{figure*}

After the data is processed in the application nodes, it is sent to the Redis Cache tier and eventually to the Amazon Relational Database Service (RDS). For the administration of the cloud application, an SSH connection is provided via Elastic Load Balancing to a Bastion server. The users need to be connected to the cloud application only on the registration process via HTTPs connection to a load balancer. The test centers send the test results using RESTful API over HTTPs. The scheduled tasks, for example, generating the user's rotational UERC, is done in an Elastic Container Service, which uses the CloudWatch event rule at a specific time of the day. The application generates alarms for any irregular activities like device connection error or access failure via CloudWatch alarms to the proper authority. Users are notified from ARC, SUDUN, and cloud application via Simple Notification Service. For further communication processes to authorized organizations and administrators, the Simple Email Service is used. For restricting unauthorized access to the applications and data layers, private subnets are used. To ensure efficient data availability in the cloud, a database instance is stored in a separate availability zone. Here, Figure~\ref{fig:Implementation Overview} illustrates the implementation overview.

\subsection{Database}
\begin{figure}[!ht]
    \centering
    \includegraphics[width=.5\textwidth]{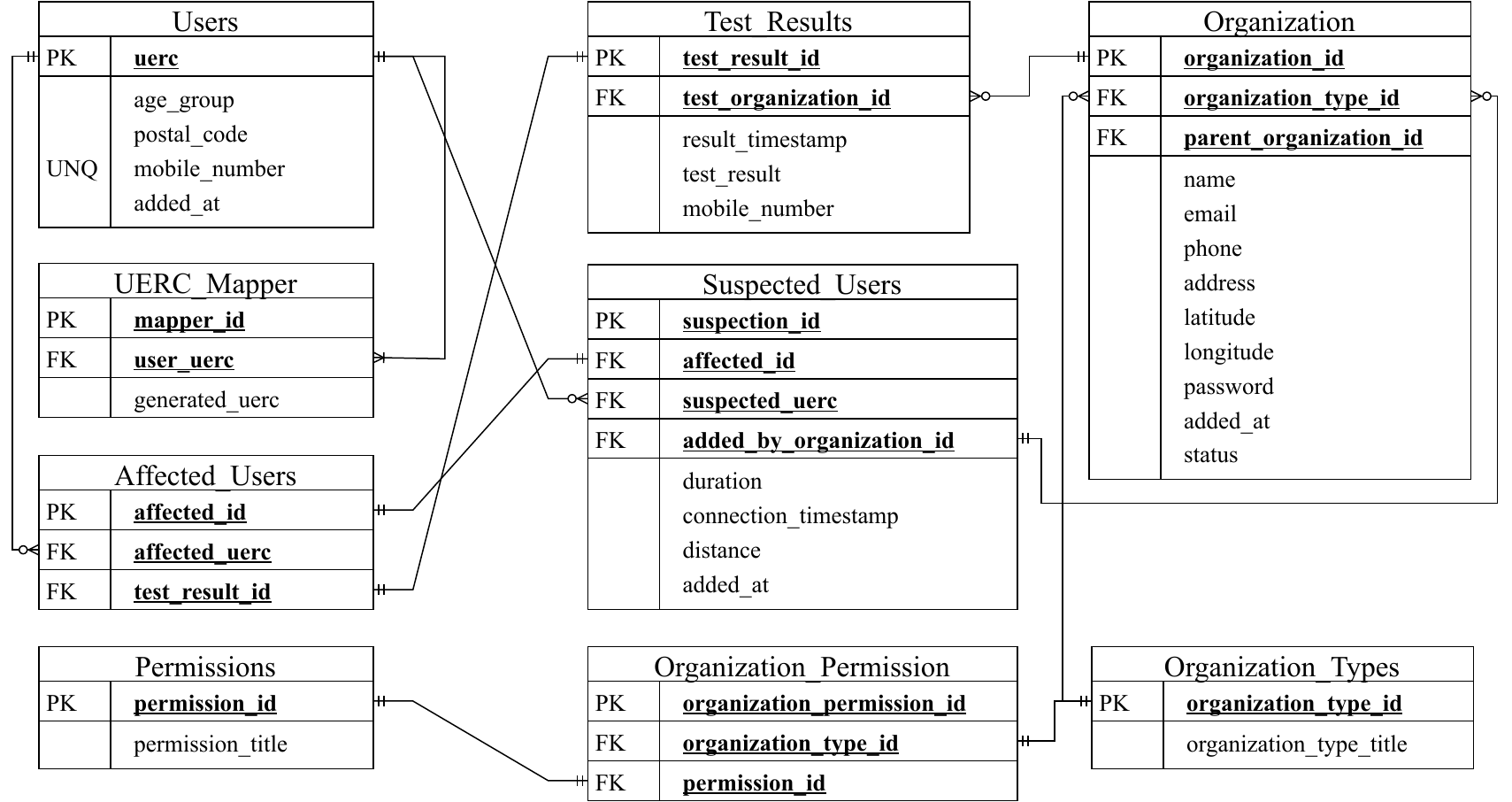}
    \caption{The cloud database consisting of mandatory tables}
    \label{fig:database_tables}
\end{figure}

The cloud database includes several tables to trace infected and suspected cases, to alert users, to prevent community transmission in organizations through ARC, and collect test results from SUDUN. The user table stores user registration information (mobile number, age group, postcode, and timestamp) and the initial UERC. As we are generating new UERCs associated with the primary user UERC, we need mapping in between those unique reference codes. We store test results (result ID, test organization ID, timestamp, test result, and user mobile number) in a separate table and extract the infected users (affected UERC, test result ID, and affected ID). Now, we map between the registered user table and affected user table to find out the suspected users and store the suspected UERC, duration, timestamp, and distance radius alongside generating a new suspect ID. We also need an organization database where we store organization information (name, email, geo-location, address.) and associated permissions for ARC and SUDUN. Figure~\ref{fig:database_tables} presents the relations between all these tables.

\subsection{Contact Tracing Graph}
\begin{figure*}
    \centering
    \includegraphics[width=.94\textwidth]{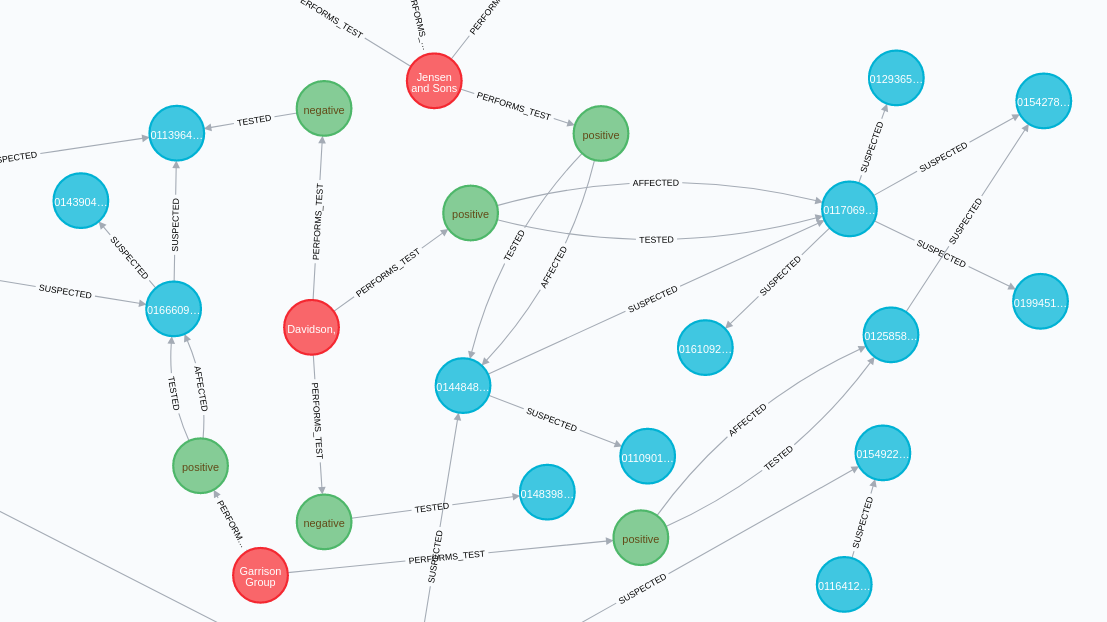}
    \caption{Contact tracing graph}
    \label{fig:contact_tracing_graph}
\end{figure*}

To identify and trace COVID-19 community transmission, we introduce and utilize a graph database called \snippet{Neo4j}~\cite{Neo4jGra88:online} to determine the contact graph. 
Figure~\ref{fig:contact_tracing_graph} presents a portion of the contact tracing graph with synthetic data of contact tracing. The red, blue, and green circles indicate the test centers, users, and test results. The lines present directed edges of the graph between infected and suspected victims.
The tracing graph can be used to identify individuals who had come in close contact with infected victims almost in real-time. We can determine a potential super spreader as well, using the contact tracing graph. Additionally, suspected victims identified by the graph can be informed to take cautionary actions to prevent community transmission.

\subsection{Privacy Preservation}
As we discussed five major data privacy issues earlier in Section~\ref{sec:background}, here, we discuss solutions to these issues in terms of implementing our framework.

\subsubsection{Voluntary} 
Users have full control over the usage of this application and the stored data in the application storage. Even it is an e-government framework, no one can force users to use the mobile app, and the participation is voluntary. While using the mobile application, users have a privacy dashboard (Figure~\ref{fig:user_privacy_dashboard}) where he can allow if the app wants to store collected RUERCs in the local storage, use device Bluetooth to share user's RUERC, and if the user wants to share the collected RUERCs with the government.


\subsubsection{Minimal Data Collection}
\begin{figure*}[t]
\centering
\begin{subfigure}{.37\textwidth}
  \centering
  \includegraphics[width=\linewidth]{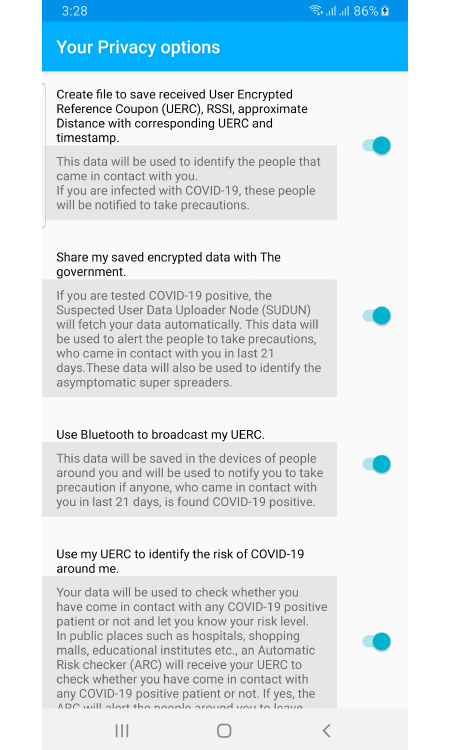}  
  \caption{User Consent Privacy Dashboard}
  \label{fig:user_privacy_dashboard}
\end{subfigure}
\begin{subfigure}{.37\textwidth}
  \centering
  \includegraphics[width=\linewidth]{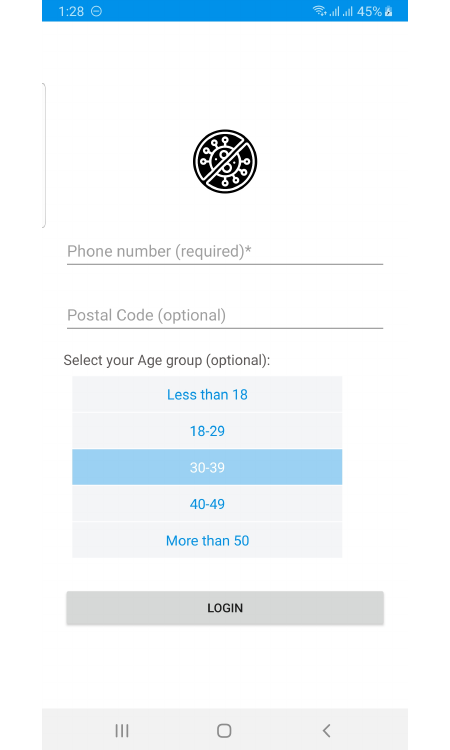}  
  \caption{Minimal Data Collection during Registration}
  \label{fig:registration_minimal_data}
\end{subfigure}
\caption{Privacy Concerned Mobile Application User Interfaces}
\label{fig:fig}
\end{figure*}

Initially, the system collects the user's \emph{age group}, \emph{postal code}, and \emph{mobile number} for registration and, in return, provides a unique ID (UERC) to be stored in a user's mobile device. Then the mobile application collects \emph{RUERC}, \emph{timestamp}, \emph{duration}, and \emph{distance} measure (based on RSSI) from other mobile devices that came in contact with the user's device. Age group and postal code are optional fields. These fields are used to define clusters and identifying super spreaders. The phone number is used to alert contact to take precautions. No personal information of the user is collected during or after the registration procedure. Figure~\ref{fig:registration_minimal_data} presents the mobile application user interface with mandatory and non-mandatory fields.


\subsubsection{Data Destruction}
The mobile application deletes collected RUERCs no later than $21$ days calculated from the latest timestamp. In the cloud application, UERC and associated data from the suspected table is deleted in $30$ days of entry timestamp. Additionally, a user has full control over his data and can delete it manually from the application storage instantly. A user can also request for account deactivation and deletion of data from the cloud database. However, it may take up to two weeks to synchronize data in all replica databases.

\subsubsection{Transparency}
Transparency is vital in data privacy, and governments should make the source codes open and publish the user data flow in the whole framework. Transparency motivates users to use an application and results in better management and increased public engagement. Consent and compliance can play a crucial role in presenting transparent data processes and improving decision making.

\subsubsection{Limited Data Usage}
The collected data should have limited use cases, and our framework opens the opportunity to determine a \emph{super spreader}, and \emph{clustered view} of positive cases. Additionally, the infected and suspected lists can help calculate the risk factors in terms of community transmission. However, while employing, further data usage should be minimal, and users need to have a clear idea of why and how data is being processed.

\section{Related Work}\label{sec:related}
Qiang Tang discussed a few existing privacy-preserving framework solutions that attempted to solve several user data privacy issues~\cite{tang2020privacy}. Author provided some observations and privacy solutions on Singapore's \snippet{Tracetogether}, Reichert et al.’s \snippet{MPC Solution}~\cite{cryptoeprint:2020:375}, Altuwaiyan et al.’s \snippet{Matching Solution}~\cite{altuwaiyan2018epic}, and Vaudenay's \snippet{DP3T} solution~\cite{cryptoeprint:2020:399}.

\paragraph*{Singapore's Tracetogether}
Affected users are compelled to share with the Ministry of Health the locally saved information. In our framework, this is voluntary to share any information with the system.
Additionally, there is a risk of decomposition of the app and collecting the geo-location data from the app. As we are not storing any geo-location data anywhere, this does not possess any risk of transpassing.

\paragraph*{Reichert et al.’s MPC Solution}
This solution requires the user’s smart device to be used to acquire and save geo-location data, and this location trace is shared with the Health Authority (HA)~\cite{cryptoeprint:2020:375}. This may raise a privacy issue of the users as it does not mention if they can preserve their privacy by controlling what information they want to share with the authority and what information they do not want to share. 
We provide a clear and concise privacy dashboard to the users so they can choose the information they want to share and alter their choices.
The solution does not consider the scalability of computation of extensive data set collected from the devices, and HA requires to arrange garbled circuits (a cryptographic protocol that encrypts computation) for all infected users. We take precautionary steps in fog nodes to filter the collected data. Additionally, the central application uses an auto-scaling technique and a queue service to facilitate the upcoming data.

\paragraph*{Altuwaiyan et al.’s Matching Solution}
This solution utilizes a privacy-preserving matching protocol between users along with proper distance measuring procedure~\cite{altuwaiyan2018epic}. We are using similar metrics to calculate the distance between mobile devices. On the other hand, infected users' exact location is shared with the central server, which indicates severe privacy violation. We do not use individual users’ locations and thus preserve users’ location privacy.

\paragraph*{The DP-3T Solution}
The DP-3T solution provides a privacy-aware framework while considering three design considerations: low-cost design, unlinkable design, and hybrid design. The solution takes advantage of the content delivery network and provides a cost estimation per patient. Authors discussed privacy concerns such as the social graph (social relationships between users), interaction graph (physical interaction nearby), location traceability (tracing individuals), at-risk individuals (suspected cases), positive status (infected cases), and exposed location (partial identification of places a positive case visited). They also discussed the addressable privacy concerns in terms of the three design considerations.

\paragraph*{COVIDSafe}
The app collects user name, phone number, age-range, and postcode followed by the consent of the user and sends to the server. The app stores encrypted user ID, time of contact, and Bluetooth signal strength and keeps this data for 21 days. The case can upload these data voluntarily~\cite{CovidSafePrivacyp9:online}. The government has amended the Privacy Act 1988 to prevent the misuse of these data~\cite{CovidSafePrivacyA5:online}. The user can uninstall the app to delete these data and to delete all data from the server, the user needs to wait until the pandemic is over whereas, our framework allows users to delete all data within 14 days. 


Table~\ref{tab:comparison} presents the difference between existing mobile application frameworks and our integrated mobile-fog computing framework (PPMF). Here, we categorize different features in terms of privacy-preserving approaches (voluntary, data usage limitation, data destruction, minimal data collection, and transparency), fog-based integrated solutions (risk check, infected/suspected data upload), and general design approaches (temporary BLE IDs, no geo-location trace, and minimal internet requirements). 
We put our framework (PPMF) at the end of the list and find it ticks all these features.

\begin{table*}[!t]
\caption{Comparison between Existing Solutions and Our Solution}
\label{tab:comparison}
\resizebox{\textwidth}{!}{ 
\begin{tabular}{|c|c|c|c|c|c|c|c|c|c|c|}
\hline
 \multirow{2}{*}{\diagbox[innerwidth = 8em,
            height = 9.5ex]{\textbf{Frameworks}}{\textbf{Features}}} & \multicolumn{5}{c|}{\textbf{Privacy-preserving Approaches}}   & \multicolumn{2}{c|}{\textbf{Fog Computing}} & \multicolumn{3}{c|}{\textbf{Design Approaches}}  \\ \cline{2-11}
 & Volunatary & \begin{tabular}[c]{@{}c@{}}Data Usage\\ Limitation\end{tabular} & \begin{tabular}[c]{@{}c@{}}Data\\ Destruction\end{tabular} & \begin{tabular}[c]{@{}c@{}}Minimal Data\\ Collection\end{tabular} & Transparency & \begin{tabular}[c]{@{}c@{}}Risk\\Check\end{tabular} &  \begin{tabular}[c]{@{}c@{}}Infected/Suspected\\Data Upload\end{tabular} & \begin{tabular}[c]{@{}c@{}}Temporary\\BLE ID\end{tabular} & \begin{tabular}[c]{@{}c@{}}No Location\\Tracking\end{tabular} & \multicolumn{1}{c|}{\begin{tabular}[c]{@{}c@{}}Minimal\\Internet Use\end{tabular}} \\ \hline\hline
 Tracetogether &  \checkmark  &  \checkmark  & \checkmark  &  \checkmark  &  \checkmark  &
 \textbf{x}   &  \textbf{x}      &       \checkmark       &      \checkmark                                                  &   \checkmark \\ \hline
 MPC  Solution & \checkmark & \checkmark & \textbf{x} & \checkmark & \textbf{x} &  \textbf{x} & \textbf{x} & \checkmark & \checkmark & \textbf{x} \\ 
 \hline
 Matching  Solution &  \textbf{x}  &  \checkmark   &  \textbf{x}   &  \textbf{x}  & \textbf{x}  & \textbf{x}  & \textbf{x}  &  \checkmark & \checkmark  & \checkmark   \\ \hline
 DP-3T  Solution &  \checkmark  &  \checkmark  &  \checkmark   &  \checkmark  & \checkmark   &  \textbf{x}  &  \textbf{x}  &  \checkmark   & \checkmark & N/A \\ \hline
 COVIDSafe & \checkmark  & \checkmark  & \checkmark & \checkmark  &  \textbf{x}   &  \textbf{x}  &  \textbf{x}  & \textbf{x}  &     \checkmark &  \checkmark  \\ \hline
 PPMF &  \checkmark  &  \checkmark  & \checkmark  &  \checkmark  &  \checkmark &  \checkmark  & \checkmark  &  \checkmark & \checkmark  &  \checkmark   \\ \hline
\end{tabular}
}
\end{table*}

\balance

\section{Conclusions}\label{sec:conclude}
In this work, we have presented a mobile and fog computing-based integrated framework that can trace and prevent community transmissions alongside maintaining user data privacy. In \emph{PPMF}, we consider minimal data collection and provide a temporary RUERC in encrypted form for storing in user devices. 
The \emph{ARC} is responsible for tracing positive cases in public places and sending alerts to nearby people without revealing the user's identity. The \emph{SUDUN} uploads data to the central database about infected and suspected cases that can be processed to identify a super spreader and visualize the clusters of cases based on postal codes and age groups.

Minimal and undetectable data collection, user control, and system transparency are essential factors to ensure user data privacy. Privacy dashboard-based on compliance and user consent makes it convenient and encourages citizens to use such a user-friendly e-government framework. Therefore, using the structure of our proposed \emph{PPMF} framework, governments can continue their economic activities while tracing and minimizing the mass-level community transmission. In the future, we plan to develop a super spreader detection model and clustering methodology of infected cases, based on our  framework.

\section*{Acknowledgement}
This research is partly supported through the Australian Research Council Discovery Project: DP190100314, 'Re-Engineering Enterprise Systems for Microservices in the Cloud.'

%
\IEEEpeerreviewmaketitle


\ifCLASSOPTIONcaptionsoff
  \newpage
\fi

%
%
\bibliographystyle{IEEEtran}
\bibliography{main}

%
%

\end{document}